\documentclass[sigconf]{acmart}

\AtBeginDocument{%
  }

\copyrightyear{2025}
\acmYear{2025}
\setcopyright{rightsretained}
\acmConference[ASPDAC '25]{30th Asia and South Pacific Design Automation Conference}{January 20--23, 2025}{Tokyo, Japan}
\acmBooktitle{30th Asia and South Pacific Design Automation Conference (ASPDAC '25), January 20--23, 2025, Tokyo, Japan}\acmDOI{10.1145/3658617.3697777}
\acmISBN{979-8-4007-0635-6/25/01}

\usepackage{subcaption}
\usepackage{amsmath,amsfonts}
\usepackage{algorithmic}
\usepackage{algorithm}
\usepackage{comment}
\usepackage{xspace}

\begin{document}

\newcommand{\alg}{{IC-D\textsuperscript{2}S}\xspace}

\title{\alg: A Hybrid Ising-Classical-Machines Data-Driven QUBO Solver Method}

\author{Armin Abdollahi}
\email{arminabd@usc.edu}
\orcid{0009-0007-1387-0995}
\affiliation{%
  \institution{University of Southern California}
  \city{Los Angeles}
  \state{CA}
  \country{USA}
}

\author{Mehdi Kamal}
\email{mehdi.kamal@usc.edu}
\orcid{0000-0001-7098-6440}
\affiliation{%
  \institution{University of Southern California}
  \city{Los Angeles}
  \state{CA}
  \country{USA}
}

\author{Massoud Pedram}
\email{pedram@usc.edu}
\orcid{0000-0002-2677-7307}
\affiliation{%
  \institution{University of Southern California}
  \city{Los Angeles}
  \state{CA}
  \country{USA}
}
\renewcommand{\shortauthors}{Abdollahi et al.}

\begin{abstract}
\vspace{-2pt}
We present a heuristic algorithm designed to solve Quadratic Unconstrained Binary Optimization (QUBO) problems efficiently. The algorithm, referred to as \alg, leverages a hybrid approach using Ising and classical machines to address very large problem sizes. Considering the practical limitation on the size of the Ising machine (IM), our algorithm partitions the QUBO problem into a collection of QUBO subproblems (called subQUBOs) and utilizes the IM to solve each subQUBO.
Our proposed heuristic algorithm uses a set of control parameters to generate the subQUBOs and explore the search space. Also, it utilizes an annealer based on cosine waveform and applies a mutation operator at each step of the search to diversify the solution space and facilitate the process of finding the global minimum of the problem.
We have evaluated the effectiveness of our \alg\, algorithm on three large-sized problem sets and compared its efficiency in finding the (near-)optimal solution with three QUBO solvers. One of the solvers is a software-based algorithm (D\textsuperscript{2}TS), while the other one (D-Wave) employs a similar approach to ours, utilizing both classical and Ising machines.
The results demonstrate that for large-sized problems ($\ge 5000$) the proposed algorithm identifies superior solutions. Additionally, for smaller-sized problems ($= 2500$), \alg\, efficiently finds the optimal solution in a significantly faster manner.
\vspace{-10pt}
\end{abstract}



\maketitle

\section{Introduction}
\vspace{-2pt}
Quadratic Unconstrained Binary Optimization (QUBO) problems involve optimizing a quadratic objective function with binary variables, making them relevant to many real-world scenarios. As NP-complete problems, they lack efficient polynomial-time algorithms for exact solutions \cite{ref23}. Many NP-complete problems, such as the graph coloring, Traveling Salesman, exact cover, and knapsack problems, can be mapped to QUBO \cite{ref8}\cite{ref23}. Finding the optimal binary values to minimize or maximize the quadratic objective function is computationally intractable, highlighting the difficulty of QUBO optimization.

Traditional methods for solving QUBO problems include branch-and-bound and stochastic optimization strategies like Tabu Search \cite{ref1, ref22}, Ant Colony \cite{ref5}, and Genetic Algorithms \cite{ref24}, all performed on conventional computing machines.

Ising machines are specialized computational devices inspired by the Ising model from statistical physics, designed to solve optimization and combinatorial problems cast as QUBO problems. These machines employ an array of interconnected nodes, typically binary elements or quantum bits (qubits), to search for optimal solutions by manipulating their nodes' states (e.g., the qubit's spin in a quantum Ising machine.) Ising machines find a configuration of nodes' states that minimizes a system's energy, making them a promising tool for tackling complex optimization challenges.
By mapping QUBO problems to the Ising formulation, the QUBO could be solved directly on the Ising machine hardware. 
Due to the applicability and efficiency of the Ising machine, hardware structures using MOS devices (e.g., \cite{ref15, ref16, ref17, ref31, ref32}), memristor-based systems (e.g., \cite{ref33}), quantum devices (e.g., \cite{ref2}), classical superconductor devices e.g., \cite{ref25}), and optical devices (e.g., \cite{ref14, ref20, ref21}) have been suggested. 

However, these machines are limited by the relatively small number of nodes (spins) that can be modeled, prompting exploring alternative strategies, such as decomposing a large QUBO problem into a plurality of sub-problems (which we call the subQUBO approach.) For example, the D-Wave quantum computer has only 5,000 nodes with a limited number of couplers \cite{ref26}, which limits the size of QUBO problems that can be directly solved on it. 
The subQUBO approach employs a hybrid framework, combining various optimization techniques with awareness about the hardware limits to navigate the complete QUBO solution space \cite{ref3}, wherein a portion of the QUBO problem is dispatched to the hardware in each iteration, with this iterative process continuing until a complete solution is obtained \cite{ref2}.

This paper presents a data-driven heuristic search algorithm (called \alg) to solve QUBO Problems on a hybrid fabric comprising Ising and conventional processing machines. Using a set of control parameters, this search algorithm uses Tabu search to generate subQUBOs to be mapped onto the Ising machine. These parameters also apply efficient mutation operators to escape local minima. The remainder of this paper is organized as follows. Section 2 will provide Background knowledge and review some related works. Section 3 explains the proposed algorithm's details. Sections 4 and 5 discuss the results and conclude the paper.

\vspace{-6pt}

\section{Background and Related Works}
\vspace{-3pt}
\subsection{QUBO Problem}

Equation (\ref{eq:1}) presents the general formulation of the QUBO problem. In this equation, the matrix $Q$ (i.e., weight matrix), sized $n\times n$, comprises integer values that encompass both positive and negative integers. The vector $x$, also of size $n$, consists of binary values, with each element $x_i$ being either one or zero. Thus, the objective is to determine the solution that minimizes the $x^T Q x$ problem, effectively solving the QUBO problem. Thus, the primary aim is to identify the vector $x$ that minimizes the function $f(x)$.

\vspace{-0.2cm}

\begin{equation}
\label{eq:1}
\small
f(x) = \sum_{i=1}^{n} Q_{ii} x_i + \sum_{i=1}^{n} \sum_{j \neq i}^{n} Q_{ij} x_i x_j\
\end{equation}

\vspace{-8pt}
\subsection{Ising formulation}
\vspace{-3pt}
Equation (\ref{eq:2}) shows the general Ising formulation, where $E$ represents the system's energy. $i$ and $j$ are indices for individual spins (nodes). $J_{ij}$ is a coupling constant determining the interaction strength between spins $i$ and $j$. It is typically a symmetric matrix describing the interactions between all spins. $s_i$ and $s_j$ are individual spin values, often taking values of $+1$ or $-1$. $h_i$ represents the strength of an external magnetic field acting on spin $i$. The transformation from Ising to QUBO is easy, and the $\{0,1\}$-valued variables of QUBO can be transformed to $\{-1,1\}$-valued variables of Ising formulation as shown in Equation (\ref{eq:3}). 

\vspace{-0.2cm}

\begin{equation}
\label{eq:2}
\small
E = -\sum_{i} \sum_{j} J_{ij} s_i s_j - \sum_{i} h_i s_i
\end{equation}

\vspace{-8pt}

\begin{equation}
\label{eq:3}
\small
s_i = 2 \times x_i - 1
\end{equation}

\vspace{-10pt}
\subsection{Related Work}
We review prior works on hybridizing classical processors and Ising machines to solve the QUBO problems. 
It should be stated that there are prior works on solving QUBO using only classical processors, such as \cite{ref1} and \cite{ref5}. 
In \cite{ref1}, a search algorithm denoted as D\textsuperscript{2}TS, has been proposed. This paper proposes a fast and efficient modified Tabu search while utilizing a unique perturbation-based diversification strategy. It shows promising results in solving large-size QUBO problems on a classical machine with few search epochs.

In \cite{ref2}, a subQUBO approach was presented to solve the QUBO optimization problem. This approach uses multiple Quantum annealer hardware to solve the subQUBO problems in combination with a Tabu search and a randomizer to find the solution for the QUBO. 
Another subQUBO approach for solving the Trip Planning Problems (TPP) using Ising machines has been introduced in \cite{ref3}. By partitioning the original QUBO model into smaller subQUBOs through combined variable selection, this paper finds near-optimal solutions to the QUBO problem. 

The authors of \cite{ref4} introduced a hybrid annealing method called subQUBO model extraction for solving QUBO models with Ising machines. This method successfully obtained multiple quasi-optimal solutions by leveraging a classical computer to identify and extract size-limited subQUBO models from selected solution instances based on variable variance. The iterative refinement of the solution set was a key aspect of this approach.
In \cite{ref10}, an algorithm for automatically planning touristic trips was introduced. This algorithm utilized Tabu search as a meta-heuristic and defined a specific set of operators to navigate the search space efficiently.
In \cite{ref28}, a metaheuristic is used in tandem with using the D-wave system quantum annealer to solve the QUBO problems. Finally, note that the Variational Quantum Eigensolver (VQE) method for hybrid usage of the classical and quantum machines for solving large-size problems has been proposed \cite{ref29}. While our proposed algorithm could be applied to any Ising machine platform (digital, analog, photonic, or quantum).

\vspace{-8pt}
\subsection{\textit{Tabu} search}
Tabu search is an optimization algorithm that explores solution spaces incrementally by modifying a current solution while avoiding recently explored solutions. This memory mechanism prevents the algorithm from getting stuck in local optima, making it effective for complex combinatorial optimization problems. 

The algorithm begins with a vector solution $x$ and runs for a fixed number of iterations ($\alpha$). In each iteration, it identifies the element of $x$ that, when inverted, yields the most significant improvement in the objective value, with a complexity of $O(n^2)$. To enhance efficiency, a data structure storing the 1-flip move value (change in $x_i$) for every possible move was proposed in \cite{ref1}. Additionally, an efficient method for updating this data structure after each iteration was suggested, reducing the time complexity from $O(n^2)$ to $O(n)$. To avoid local minima, Tabu search incorporates a memory mechanism. Specifically, a value flipped in one iteration within the $x$ vector cannot be flipped for the subsequent $c$ iterations unless it results in a better solution. This prevents immediate undoing of beneficial moves and promotes diversity by discouraging recent moves, aiding in escaping local optima.

Algorithm \ref{alg:Tabu} depicts the pseudo-code for the Tabu search algorithm used in this paper, based on \cite{ref1}. The algorithm initiates by computing all $\Delta x_{k}$ values, representing the potential change in the objective function if the $k$\textsuperscript{th} variable is flipped. Within the main loop, the $C$ vector dictates which elements can be flipped; a positive value in the $C$ vector indicates that the corresponding variable cannot be flipped unless it becomes negative in subsequent iterations. This serves as the memory function, and the allowable elements for flipping are stored in the $filtered\_vec$ at each iteration. The variable yielding the highest difference is selected for flipping, leading to a recalculation of the $\Delta x_{k}$ values for the next iteration. The chosen element of the $C$ vector has $c$ added to it, while all values in the $C$ vector are decremented by one. The objective value and $x_{min}$ values are then updated correspondingly.

\begin{algorithm}[t]
\caption{Tabu search}
\label{alg:Tabu}
\begin{algorithmic}[1]
\setlength{\itemsep}{0.2em} 
\setlength{\baselineskip}{0.8em} 
\small
\STATE OV = $x^TQx$ $\;\;\;\;\;\;$ \small \%OV: Object Value \normalsize
\vspace{-1.5pt}
\STATE $OV_{min} = OV$
\vspace{-1.5pt}
\FOR{$k$ in $1$ to $n$}
\vspace{-1.5pt}
    \STATE $\Delta x_k = (1 - 2x_k) \left(Q_{k,k} + \sum_{j \in n; j \neq k, x_j = 1} 
    Q_{j,k}\right)$
    \vspace{-1.5pt}
\ENDFOR
\vspace{-1.5pt}
\FOR{$\alpha$ iterations}
\vspace{-1.5pt}
    \FOR {each k}
    \vspace{-1.5pt}
        \IF {$C[k] <= 0$}
        \vspace{-1.5pt}
            \STATE $filtered\_vec[k] = \Delta x_k$
            \vspace{-1.5pt}
        \ELSE
        \vspace{-1.5pt}
            \STATE $filtered\_vec[k] = INF$
            \vspace{-1.5pt}
        \ENDIF
        \vspace{-1.5pt}
    \ENDFOR    
    \vspace{-1.5pt}
    \STATE $i \gets argmin(filtered\_vec)$ 
    \vspace{-1.5pt}
    \STATE $\Delta x_i \gets -\Delta x_i$
    \vspace{-1.5pt}
    \FOR{each $i \in n; j \neq i$}
    \vspace{-1.5pt}
        \IF{$x_{j} = 1$}
        \vspace{-1.5pt}
            \STATE $\Delta x_j = \Delta x_j + Q_{i,j}$
            \vspace{-1.5pt}
        \ELSE
        \vspace{-1.5pt}
            \STATE $\Delta x_j = \Delta x_j - Q_{i,j}$
            \vspace{-1.5pt}
        \ENDIF
        \vspace{-1.5pt}
    \ENDFOR
    \vspace{-1.5pt}
    \STATE $OV = OV + \Delta x_i$
    \vspace{-1.5pt}
    \STATE $C[i] += c$
    \vspace{-1.5pt}
    \STATE $x_i = 1 - x_i$
    \vspace{-1.5pt}
    \FOR {each i}
    \vspace{-1.5pt}
        \STATE $C[i]-=1 \text{ IF } C[i] > 0$
        \vspace{-1.5pt}
    \ENDFOR
    \vspace{-1.5pt}
    \IF {$OV < OV_{min}$}
    \vspace{-1.5pt}
        \STATE $x_{min} = x$
        \vspace{-1.5pt}
        \STATE $OV_{min} = OV$
        \vspace{-1.5pt}
    \ENDIF    
    \vspace{-1.5pt}
\ENDFOR
\vspace{-1.5pt}
\RETURN $x_{min}$
\end{algorithmic}
\end{algorithm}

\begin{algorithm}[t]
\caption{Proposed \alg}\label{alg:alg2}
\begin{algorithmic}[1]
\small
\REQUIRE $Q$
\vspace{-1pt}
\ENSURE $x_{\text{min}}$ for $x^T Q x$ problem
\vspace{-1pt}
\STATE $S \gets$ Generate\_Initial\_Solution\_Set()
\vspace{-1pt}
\STATE $S \gets$ Call\_IM\_Solution\_Set($S$)
\vspace{-1pt}
\STATE $r \gets$ MRAnnealer.Initial()
\vspace{-1pt}
\STATE $\eta \gets$ Weight\_Effect\_Extraction()($Q$)
\vspace{-1pt}
\WHILE{Termination Condition is not met}
\vspace{-1pt}
    \STATE $S, \Delta \gets$ Tabusearch(Q, $\alpha$, $S$)
    \vspace{-1pt}
    \STATE $\gamma \gets$ \text{Deviation\_Extraction}($S$)
    \vspace{-1pt}
    \STATE $A \gets$ Control\_Parameters\_Aggregation($\eta, \Delta, \gamma$)
    \vspace{-1pt}
    \STATE $S, S' \gets$ Call\_IM\_Partial\_Solution\_Set($S, A, Q$)
    \vspace{-1pt}
    \STATE $S \gets$ Mutation($S, S', A, r$)
    \vspace{-1pt}
    \STATE $r \gets$ MRAnnealer.NextRate() 
    \vspace{-1pt}
\ENDWHILE
\vspace{-1pt}
\RETURN $x_{\text{min}}$
\vspace{-1pt}
\end{algorithmic}
\label{alg:proposed_heuristic}
\end{algorithm}

\vspace{-7pt}
\section{Proposed heuristic}
\vspace{-2pt}
The pseudo-code of the proposed heuristic is provided in Algorithm \ref{alg:proposed_heuristic}, which aims to find the solution ($x$) that minimizes the $x^T Q x$ value. The input of the algorithm is the matrix $Q$, and its output is the (near-)optimal solution ($x_{min}$). The algorithm initiates by generating a set of random solutions, denoted as $S$, comprising a binary matrix with $z$ rows and $n$ columns (line 1). Subsequently, the solutions within matrix $S$ are updated using the Ising machine (IM). Each random solution in $S$ is divided into non-overlapping smaller solutions of size $m$, where $m$ represents the Ising machine size. For instance, the solution in the $k^{th}$ row of $S$ is segmented into $\lfloor n / m \rfloor$ subsolutions, labeled as $S_{p,l}$, where $p$ ($l$) indicates the solution (subsolution) index. In the following steps, each subsolution is substituted with the solution extracted by the Ising machine for the relevant variables of that subsolution. Simultaneously, the influence of the remaining subsolutions (i.e., the values of other variables) is factored in as a bias ($b$) for the coefficients in the linear part of the QUBO formulation.
In this case, the subQUBO formulation for the variables of $S_{p,l}$ is defined by
\vspace{-6pt}

\begin{equation}
\small
\begin{aligned}
f(S_{p,l}) & = \sum_{i \in S_{p,l}} (Q_{i,i} + b_i) x_i + \sum_{i, j \in S_{p,l}} Q_{i,j} x_i x_j \\
b_i & = \sum_{j \in S_p, j \notin S_{p,l}} (2.Q_{i,j})x_j
\end{aligned}
\label{eq:4}
\end{equation}

\begin{figure}[ht]
\centering
\includegraphics[width=2.2in]{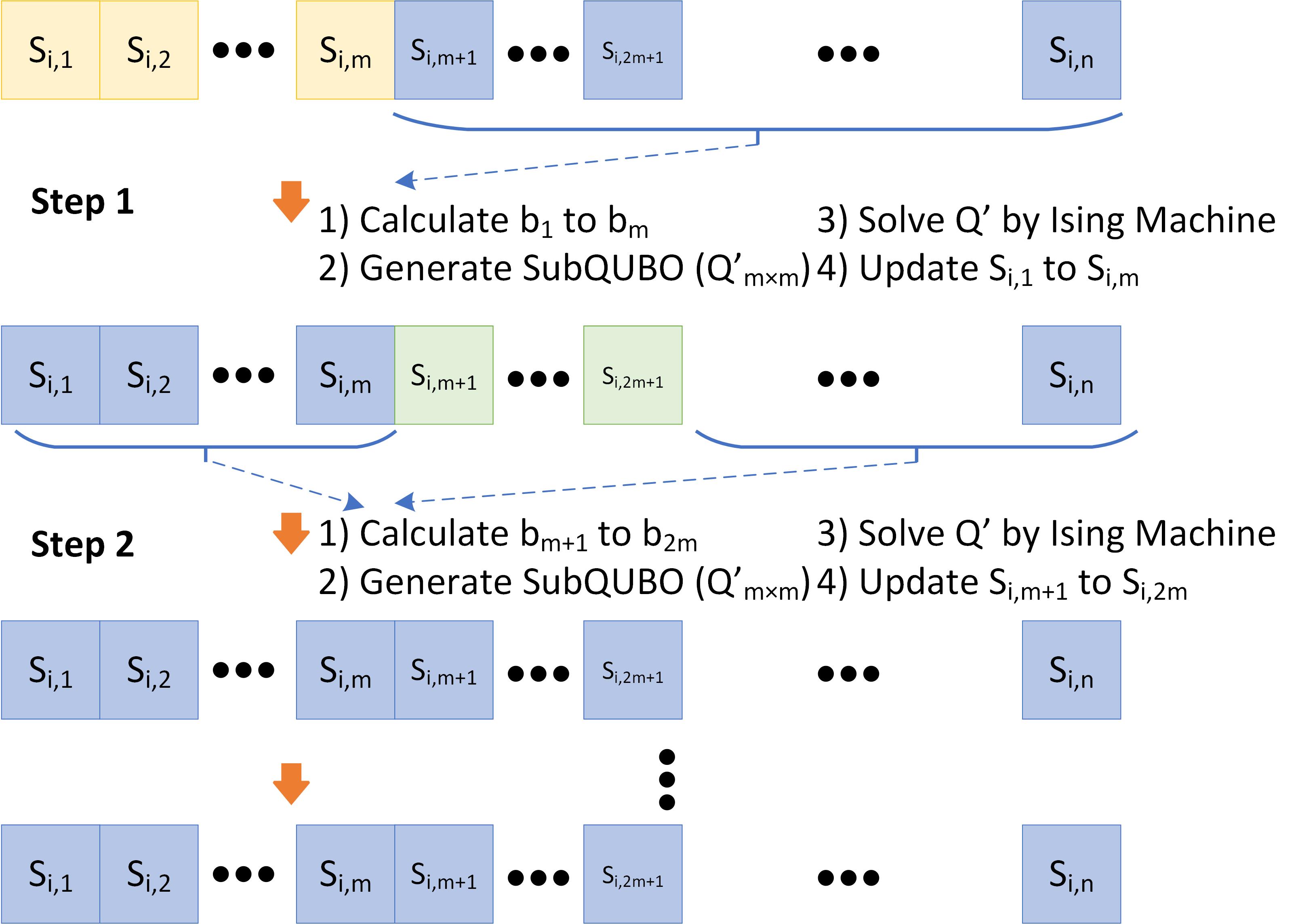}
\caption{\small \text{The procedure of \textit{Call\_IM\_Solution\_Set()} fucntion}}
\label{fig:IM_Solutionset}
\end{figure}

Upon obtaining the precise solution for a subQUBO, the respective subsolution within the solution set $S$ is revised, incorporating the updated values for extracting the bias of the subsequent QUBO subproblem.
It is essential to highlight that in scenarios involving multiple Ising machines, several subQUBOs can be addressed concurrently, thereby enhancing the efficiency of this algorithmic segment. This process is accomplished utilizing the \textit{Call\_IM\_Solution\_Set()} function (line 2). As an example, the process of updating two subsolutions of $S_i$ is shown in Fig. \ref{fig:IM_Solutionset}.

In the proposed search algorithm for escaping from local minimum points, we use a mutation mechanism, whose details will be described at the end of this section. The mutation rate is controlled by a Cosine Annealer (denoted by MRAnealer, see line 3 of the pseudo-code.) 
This parameter is adjusted after each search loop in the proposed algorithm. Its gradual convergence mechanism ensures thorough exploration of solution space while mitigating the risk of premature convergence to suboptimal solutions. 
The cosine Annelear in t\textsuperscript{th} iteration is formulated by 
\begin{align}
\small
    r_t = 0.3 \times \left(1 + \cos\left(\frac{\pi \cdot t}{15}\right)\right) \times \left(0.99^t\right) 
\end{align}

\vspace{-10pt}

\begin{figure}[!t]
\centering
\includegraphics[width=2.0in]{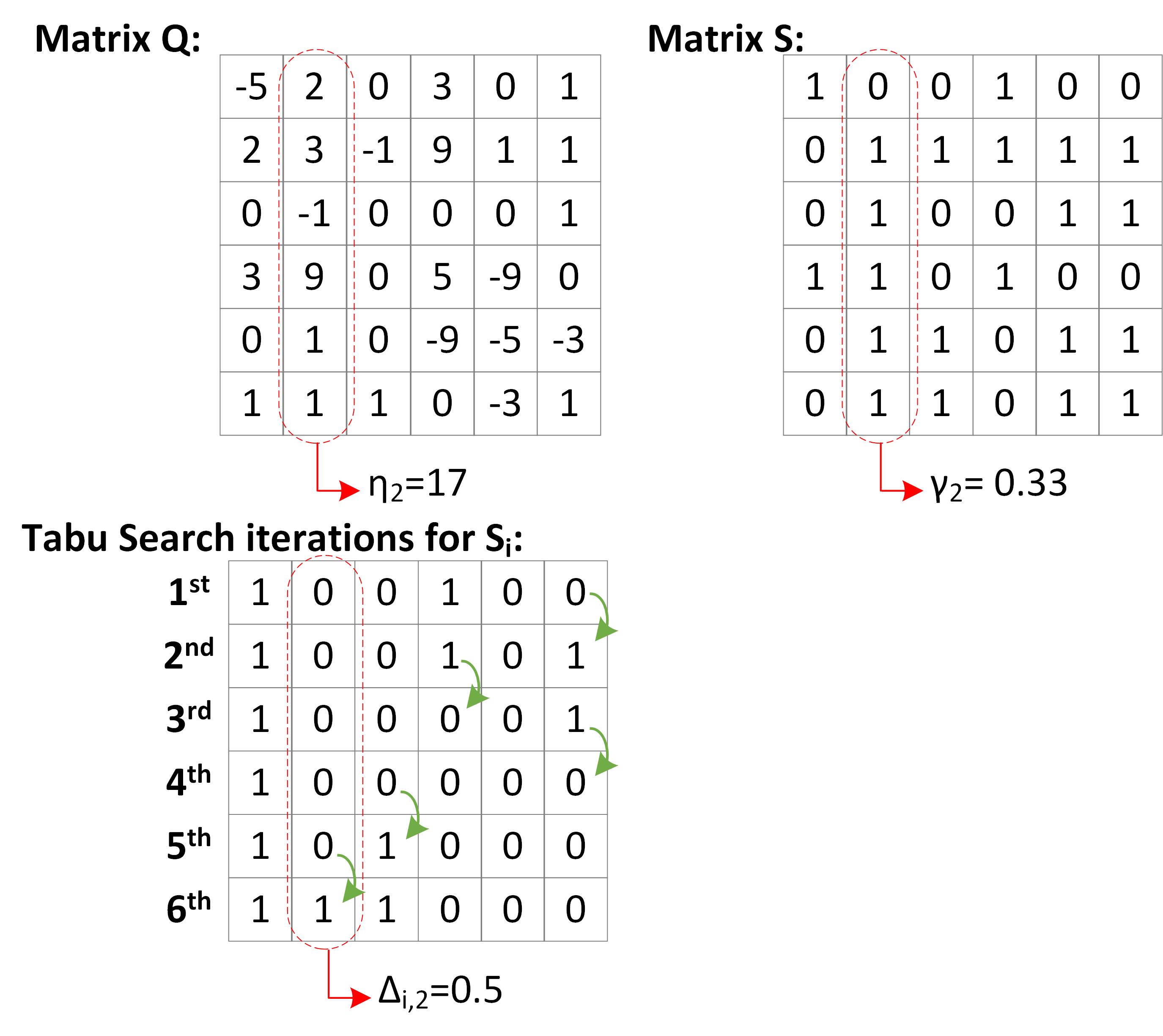}
\caption{\small Extracting the control parameters.}
\label{fig:impactparameters}
\vspace{-2em}
\end{figure}

In \alg, we introduce three control parameters ($\eta$, $\Delta$, and $\gamma$) that govern the search direction within the search space. Fig. \ref{fig:impactparameters} illustrates three instances demonstrating the extraction of these parameters. Among these parameters, $\eta$ is established once before the algorithm's main loop, while the remaining parameters are updated at each iteration. 

The parameter $\eta$ is a vector of length $n$, representing the influence of each variable on modifying the output of $f(x)$ based on the inter-variable weights. Higher weight values indicate a more pronounced impact on altering the output value. Therefore, this parameter encapsulates the interrelations among the matrix elements, proving instrumental in identifying the global minimum. The computation of this parameter is carried out using the \textit{Weight\_Effect\_Extraction()} function in line 4 of Algorithm \ref{alg:proposed_heuristic}. The j\textsuperscript{th} element of this vector ($\eta_j$) is calculated as

\vspace{-5pt}
\begin{equation}
\label{eq:if_eta}
\small
\eta_{j} = \sum_{i=1}^{n} \left| Q_{i,j} \right|
\end{equation}

Within the main loop of the suggested algorithm, we initially execute the \textit{Tabu} search on the existing solutions in $S$. The primary iteration of the Tabu search (refer to Algorithm \ref{alg:Tabu}) is reiterated $\alpha$ times. Upon completing the Tabu search, we derive the second control parameter ($\Delta$). This parameter comprises a $(z \times n)$ matrix reflecting the stability rate of each element in $S$ throughout the Tabu search phase. To achieve this, we monitor the toggling count of each element denoted by the parameter $T$. $T_{i,j}$ is the total flips the $j^{th}$ element of solution i is experiencing in the Tabu search. 
 Thus, the $\Delta_{i,j}$ parameter is determined by

\vspace{-0.4cm}

\begin{equation}
\label{eq:if_Delta}
\small
 \Delta_{i,j}= 1 - \frac{T_{i,j}}{{\max\left\{T_i\right\}}}
\end{equation}

A higher value of a $\Delta$ element indicates that the algorithm is more likely to find the (near-)optimal value for the corresponding variable.   
In the next step, parameter $\gamma$ is calculated by calling \textit{Deviation\_Extraction()} function (line 7). This parameter is a vector of length $n$ showing the deviation of each variable in the solution set of $S$. The j\textsuperscript{th} element of this vector ($\gamma_j$) is obtained by

\vspace{-0.3cm}

\begin{equation}
\label{eq:if_gamma}
\small
\mathrm{\gamma}_{j} = 1 - \frac{|\left(\sum_{i} S_{i,j} - \frac{z}{2}\right)|}{\left(\frac{z}{2}\right)}
\end{equation}

The reason behind considering this parameter is that due to the size limit of the IM, we should use the machine to determine values of the variables with higher fluctuation (i.e., high $\gamma$ values) after the Tabu search. A higher value of this parameter for a variable shows the Tabu search will require more search to find the optimal value for that variable. 

After extracting the control parameters, they are aggregated by calling \textit{Control\_Parameters\_Aggregation()} function (Algorithm \ref{alg4}). We assign a weight (\textit{w}) to aggregate the parameters. 
The weight values are set heuristically.
The output of this function is a matrix (denoted by \textit{A}), each element of which shows the importance of determining the value of the corresponding variable by using the IM. 
Thus, in the next step (line 9), we call the IM machine to solve a portion of each solution in $S$. 
The set of variables in the subQUBO problem that will be mapped on the IM corresponds to variables with the highest values in $A$. The Ising machine is called just once for each solution of $S$. 
Note that when we have a multi-IM fabric, we generate a set of subQUBOs for each solution of $S$ to be solved on the plurality of Ising machines in parallel. 
Algorithm \ref{alg3} shows the details of generating the subQUBO and solving it by IM.

\begin{algorithm}[H]
\caption{Control\_Parameters\_Aggregation}\label{alg:alg4}
\begin{algorithmic}[1]
\small
\vspace{-1.5pt}
\FOR{$i < z$}
\vspace{-1pt}
    \STATE $A_i \gets w_1 \cdot \eta  + w_2 \cdot \gamma - w_3 \cdot \Delta_i$
    \vspace{-1.5pt}
\ENDFOR
\vspace{-1.5pt}
\RETURN $A$
\vspace{-1.5pt}
\end{algorithmic}
\label{alg4}
\end{algorithm}

\vspace{-0.7cm}

\begin{algorithm}[H]
\caption{Call\_IM\_Partial\_Solution\_Set}\label{alg:alg3}
\begin{algorithmic}[1]
\small
\FOR{$p < z$}
\vspace{-1pt}
    \STATE $S'_{p}, \gets$ Subset of $S_{p}$ that has the highest $A_{p}$ values
    \vspace{-1pt}
    \STATE $Q' \gets$ Submatrix of Q (subQUBO) based on the variables in $S'_{p}$
    \vspace{-1pt}
    \STATE $b \gets$ Calculate by Equation (\ref{eq:4}) based on variables of $S'_{p}$
    \vspace{-1pt}
    \STATE $Q' \gets$ Add elements of $b$ with elements on the diagonal
 of $Q'$
 \vspace{-1pt}
    \STATE $S'_{p} \gets$ Solve subQUBO on Ising machine
    \vspace{-1pt}
    \STATE $S_{p} \gets$ Update the values of the corresponding variables based on the values of $S'_p$
    \vspace{-1pt}
\ENDFOR
\vspace{-1pt}
\RETURN $S, S'$
\vspace{-1pt}
\end{algorithmic}
\label{alg3}
\end{algorithm}

After fine-tuning the solutions by solving the corresponding subQUBO problems, the proposed heuristic applies the mutation operation (line 10) to a subset of variables in each solution of $S$. This subset does not include the variables that the IM has solved in \textit{Call\_IM\_Partial\_Solution\_Set()}. Thus, each solution's subset contains the variables with the highest values in $A$. Note that the size of each subset for each solution is  $(n-m) \times r$. 
After identifying the subset, the probability of flipping each variable of the subset is obtained based on the normalized value of the corresponding value of the variable in $A$. 

In the last step of the main loop, the annealer updates the rate of the mutations for the next iteration. 
The main loop of the algorithm iterates until a termination condition is satisfied.
We set the termination condition as the number of consecutive iterations where the best solution remains unchanged. 
Finally, the best solution for $S$ is thus computed and returned by the proposed heuristic.

\vspace{-6pt}

\section{Results}
\vspace{-3pt}
\subsection{Experiment Setup}
\vspace{-2pt}
The proposed algorithm was implemented in Python and executed on a system with an Intel(R) Core(TM) i7-10750H CPU and 16GB RAM. To assess the heuristic's efficacy, we considered three problem sets with Q sizes of 2500 $\times$ 2500, 5000 $\times$ 5000, and 7000 $\times$ 7000. Problems with $n=2500$ are available in \cite{ref7} and are denoted as Beasley matrices (Q1 to Q10). Other matrices, known as Palubeckis matrices, are available in \cite{ref30}. Beasley matrices are sparse, real-world-inspired benchmarks for QUBO, while Palubeckis matrices are dense, randomly generated for testing algorithm robustness.

The effectiveness of our proposed algorithm is compared to the D$^{2}$TS \cite{ref1}, and D-Wave algorithm \cite{ref2} and \cite{ref28}. As mentioned before,  D$^{2}$TS is an efficient software-based algorithm for solving the QUBO, whereas the D-Wave algorithm is a hybrid algorithm (similar to ours) to solve QUBO. We have implemented these two algorithms with Python. Similar coding styles and optimization are used to implement the algorithms. Note that, as previously mentioned, there are a few hybrid algorithms (e.g., \cite{ref3, ref4}) for solving QUBO problems using both classical processors and Ising machines. However, all of them, except \cite{ref28} and D-Wave algorithm, have only been evaluated on small-size problems ($2500 \gg n$). Our results show that the algorithm in \cite{ref3, ref4} is unable to find the optimal solutions for some Beasley problems after 300 epochs. \cite{ref28} also struggles to find the optimal answer of some of the Beasley matrices, and their success rate is pretty low for Palubeckis instances. In contrast, as we will demonstrate, our proposed algorithm and the D-Wave algorithm, can find the optimal solutions for this problem set in just a few epochs.

if we assume a constant time complexity, $O(1)$, for Ising machine to solve a subQUBO problem, the time complexity of the proposed algorithm is $O(z \cdot \alpha \cdot n)$. The time complexity of D$^{2}$TS and D-Wave are also $O(\alpha \cdot n)$. Note that the $\alpha$ parameter for D$^{2}$TS and D-Wave is set to $20n$ (see \cite{ref1} and \cite{ref2}.) In all studies, the Ising machine was modeled by \cite{ref27}, using dwave\_qbsolv Python package, which has been developed by D-Wave Systems. 
Also, in all studies, we assumed the Ising machine size was 50. It is clear that by increasing the size of the machine, the convergence speed for finding the (near-)optimal solution is increased.
However, we have chosen a small size value (for the size of the considered problems) to study the efficacy of the algorithms on efficient usage of the IM machine with a considerably smaller size compared to the input problem. 
Also, we assume that only one IM machine is available for all studies conducted.

\vspace{-0.3cm}
\subsection{Impact of Hyperparameters}

The values for the hyperparameters of the proposed heuristic, including the weights for Parameter Aggregation, are provided in Table \ref{Table:1}, which we have used for comparison studies. In this subsection, we study the impact of each hyperparameter, including the weights, on the heuristic's effectiveness by sweeping their values in solving the Q2 problem of the Beasley problem set. Note that while one parameter is swept, the values of the other parameters remain constant and equal to those reported in Table \ref{Table:1}.

\vspace{-9pt}

\begin{table}[htp]
\caption{\small The default value of the hyperparameters}\label{Table:1}
\vspace{-1em}
\centering
\renewcommand{\arraystretch}{0.4}
\begin{tabular}{|c|c|c|c|c|c|c|}
\hline
$z$ & $c$ & $\alpha$ & $w_1$ & $w_2$ & $w_3$ \\
\hline
4 & n/150 & \textit{5n} & 1.0 & 1.0 & 0.5 \\
\hline
\end{tabular}
\vspace{-1em}
\end{table}

\begin{figure*}[tbp]
\vspace*{-1mm}
     \centering
     \begin{subfigure}[]{0.33\textwidth}
         \centering
         \includegraphics[width=0.92\textwidth]{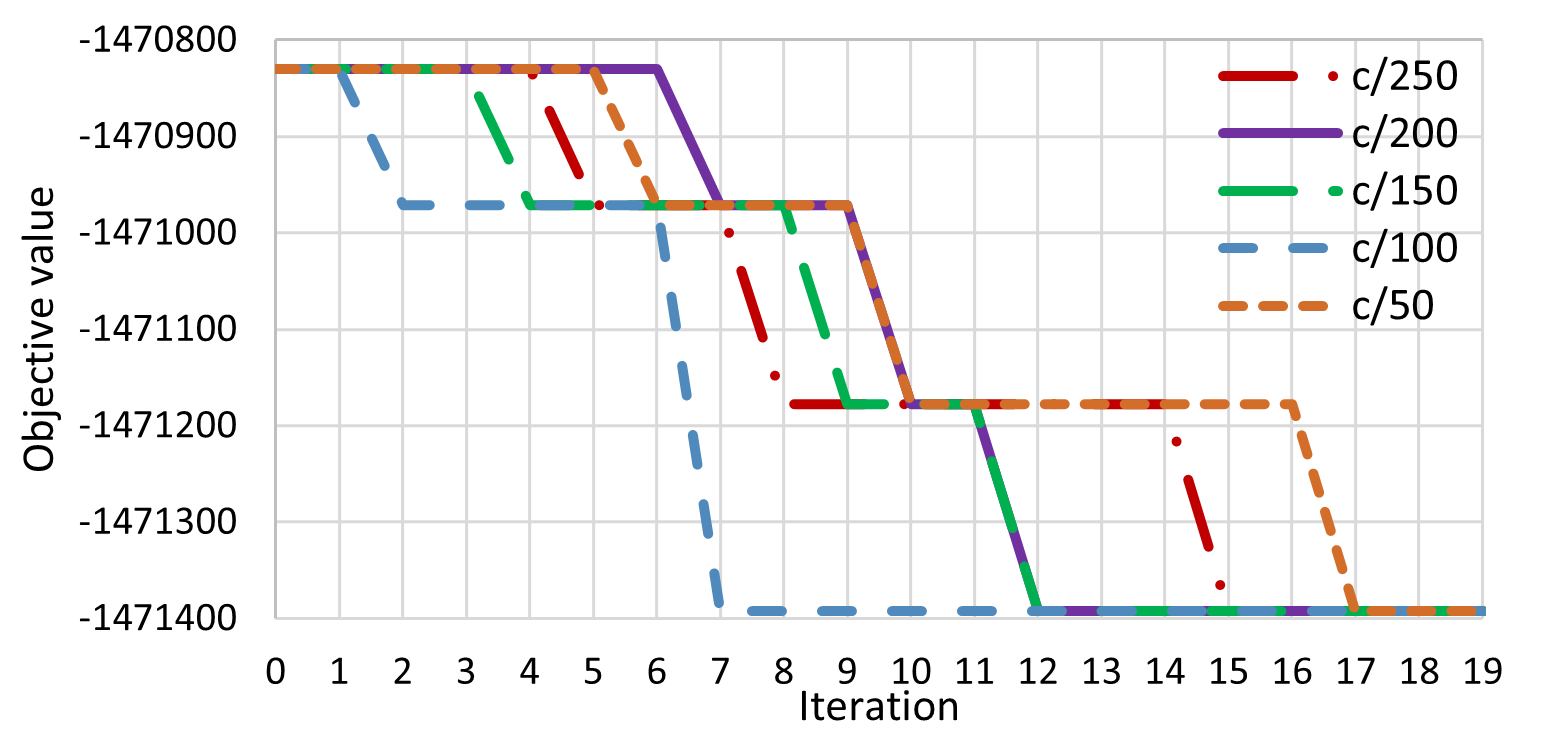}
         \caption{Impact of c}
         \label{fig:6}
     \end{subfigure}
     \hfill
     \begin{subfigure}[]{0.33\textwidth}
         \centering
         \includegraphics[width=0.96\textwidth]{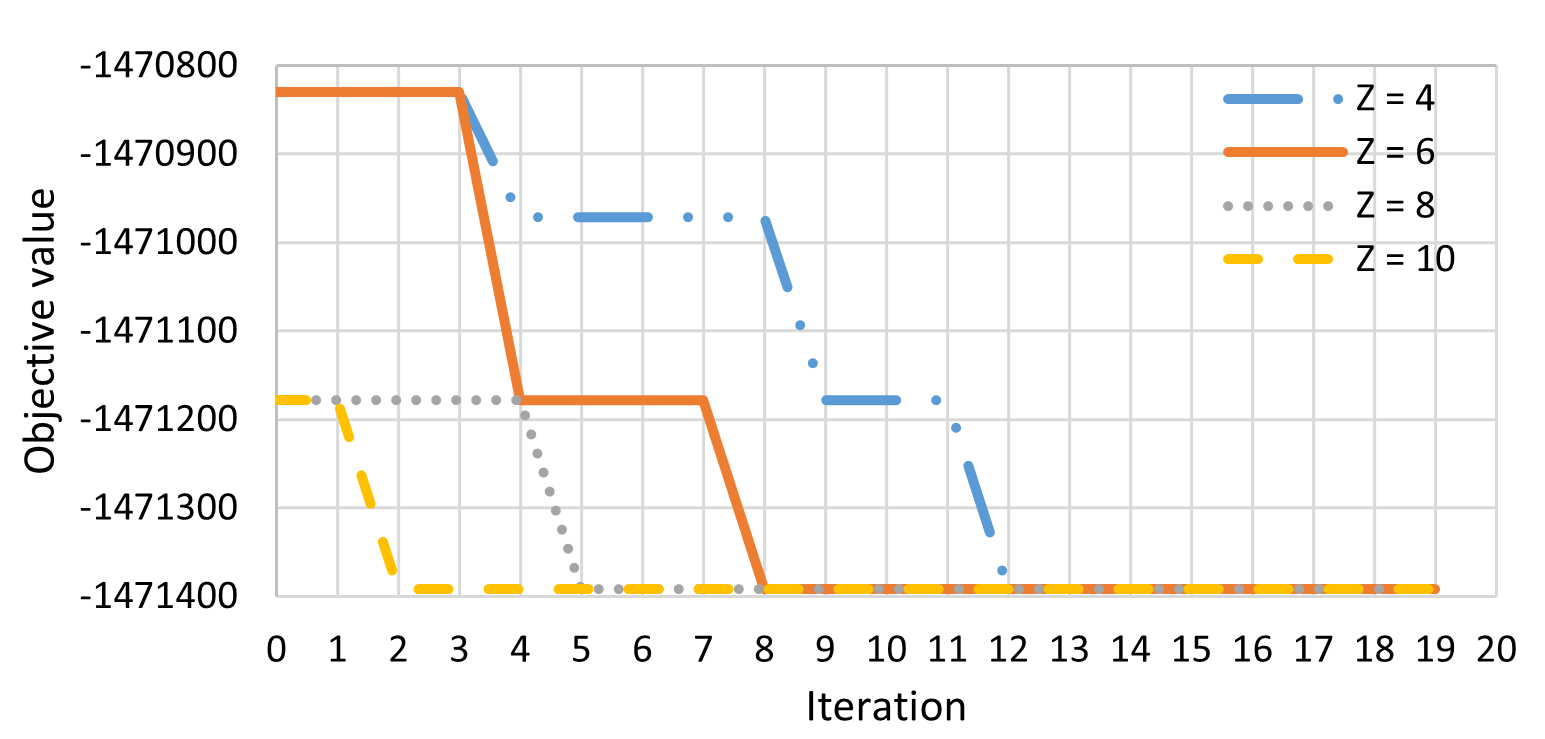}
         \caption{Impact of z}
         \label{fig:7}
     \end{subfigure}
     \hfill
     \begin{subfigure}[]{0.33\textwidth}
         \centering
         \includegraphics[width=0.92\textwidth]{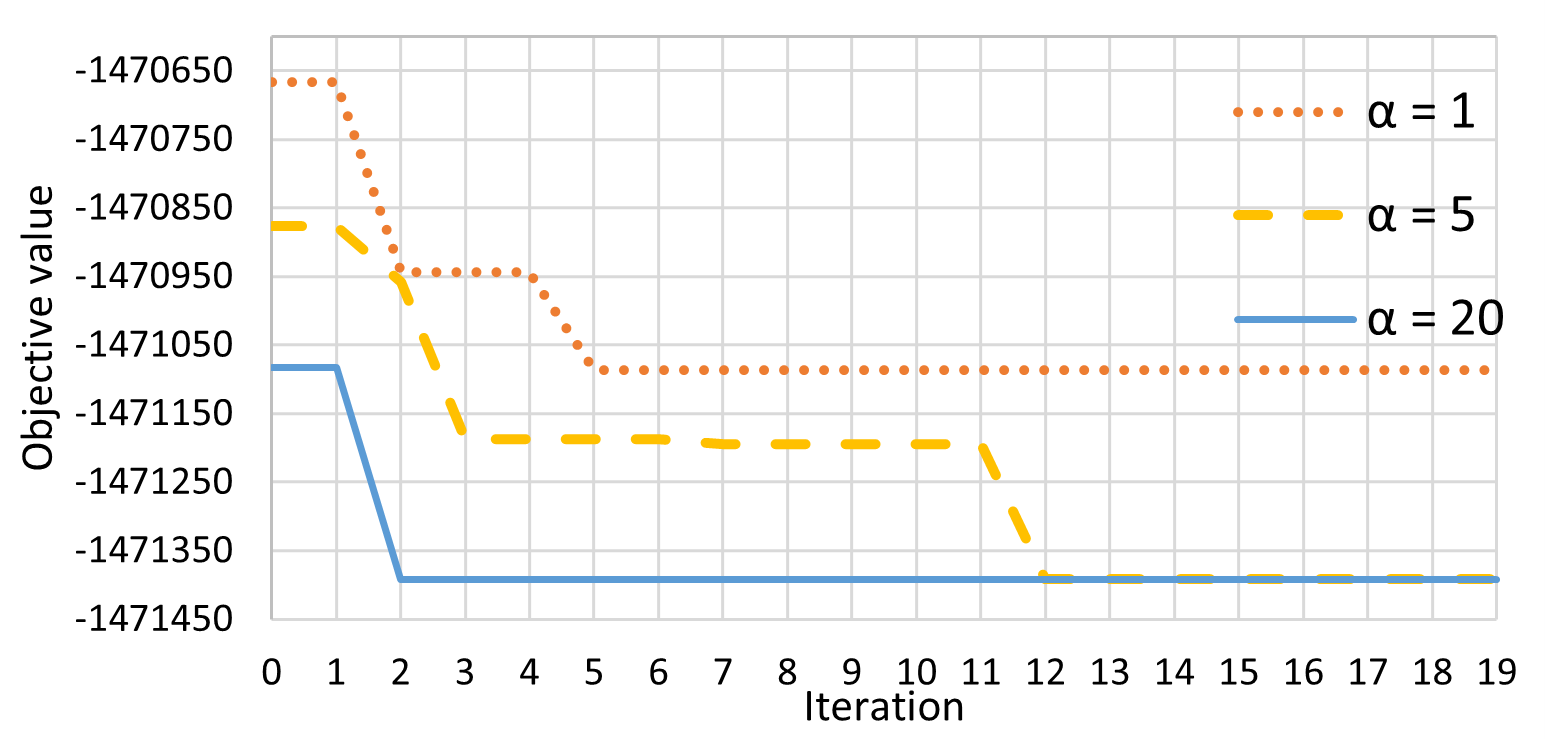}
         \caption{Impact of $\alpha$}
         \label{fig:9}
     \end{subfigure}
        \caption{\small Impact of different values of a) $c$, b) $z$ and c) $\alpha$ on the effectiveness of the proposed heuristic on solving the Q2 problem of Beasley problem set.}
        \vspace*{-2mm}
        \label{fig:6_7_9}
        \vspace{-1em}
\end{figure*}

Figure \ref{fig:6} shows the objective values of the Q2 problem under different $c$ values of the proposed heuristic. Note that $c$ is a Tabu search parameter that controls its memory mechanism. 
As the results show, the impact of this parameter is entirely unpredictable. Thus, based on our observation of some problems, we have chosen $n/150$ as the default value for this parameter. 

The effect of parameter $z$ on the efficacy of the proposed algorithm is illustrated in Figure \ref{fig:7}. Increasing the value of $z$ results in a higher number of candidate solutions being considered in each epoch. Consequently, the likelihood of discovering optimal solutions within fewer epochs is enhanced, albeit at the expense of extended runtime for each epoch. 
As shown in the figure, elevating the $z$ parameter from 4 to 10 (2.5 times larger) and specifically with $z=10$, \alg{ } managed to identify the optimal solution in three times fewer epochs compared to the scenario where $z=4$. Therefore, a higher value of $z$ could potentially lead to reduced runtime. For the sake of fair comparison in subsequent studies, we opted for $z=4$, establishing parity in complexity between our heuristic and those delineated in \cite{ref1} and \cite{ref2}.



Figure \ref{fig:9} shows the impact of parameter $\alpha$ on the quality of the proposed heuristic search algorithm. Similar to parameter $z$, increasing it improves the convergence speed and the quality of the solution at the cost of higher runtime complexity. Our studies show that $\alpha=5n$ can find the best solution. Thus, to make the runtime complexity of the proposed algorithm the same as the D$^2$TS and D-Wave algorithms (with the $\alpha=20n$), we have chosen $\alpha=5n$ that by considering the value of parameter $z$ as 4, the runtime complexity of the three algorithms is the same. 

\vspace{-8pt}

\subsection{Discussion}
\vspace{-1pt}
In this subsection, we conduct a comparative analysis of the efficiency of the proposed heuristic in contrast to the D-Wave and D$^2$TS algorithms. Each algorithm was executed ten times for each problem, with the average of the objective values recorded in each epoch. On average, our proposed heuristic required approximately 5 times fewer epochs compared to D$^2$TS (4 times fewer than D-Wave) to find the optimal solution.

\begin{figure}[tpb]
\vspace{-1em}
\centering
\includegraphics[width=2.5in]{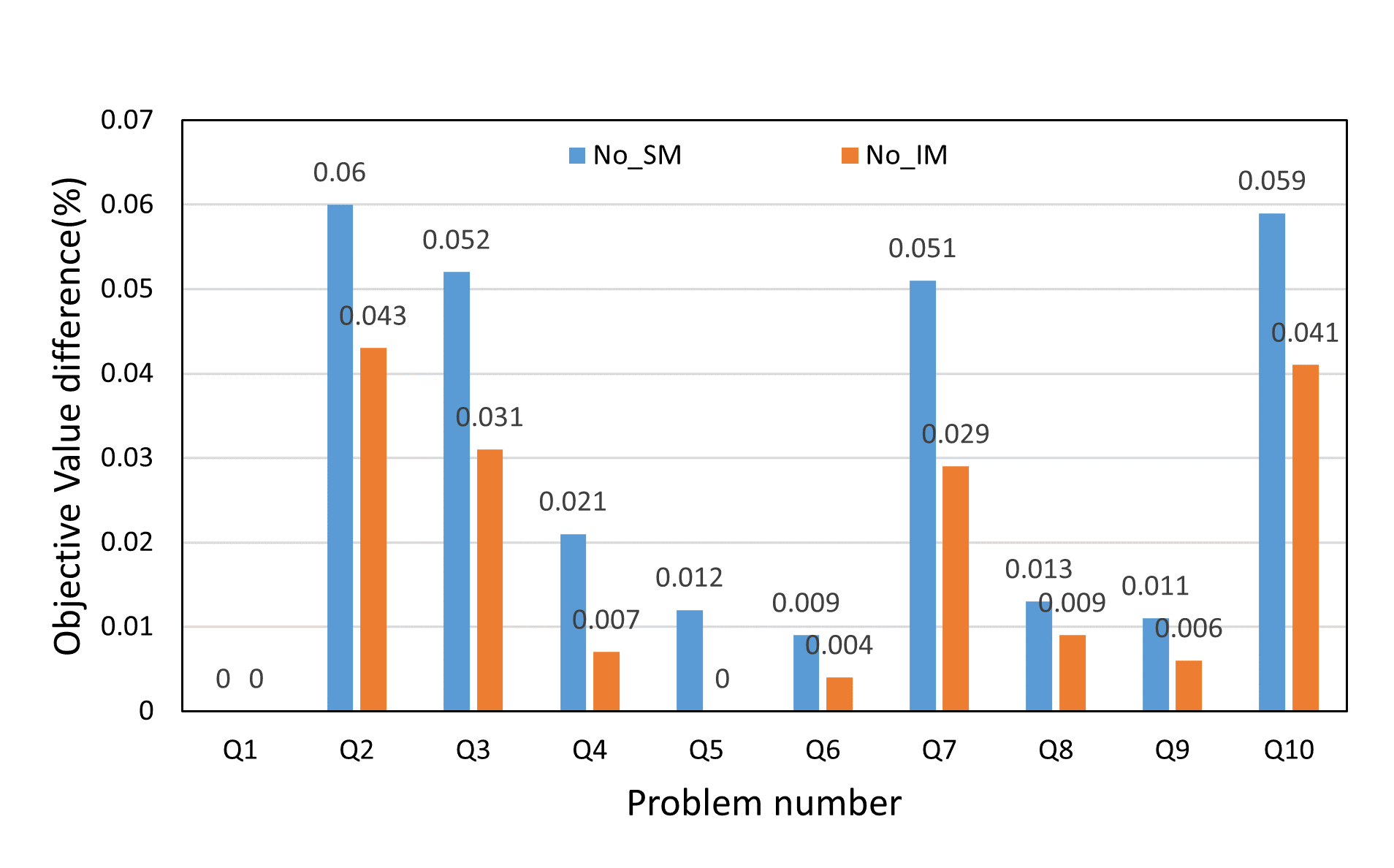}
\vspace{-1em}
\caption{\small Comparing the efficiency of the two simplified versions of the proposed heuristic compared to the original version.}
\label{fig:23}
\vspace{-1em}
\end{figure}

Despite both our algorithm and the D-Wave algorithm utilizing the Ising machine, our approach exhibits notably superior performance. This enhanced efficiency can be attributed to the innovative QUBO partitioning strategy and the utilization of specific control parameters to navigate the search space effectively.

To emphasize the effectiveness of our algorithm, we implemented two simplified versions for comparison. The first version (No\_SM) omits the proposed control parameters at various heuristic stages, while the second version (No\_IM) excludes the Ising machine. Figure \ref{fig:23} shows the difference in objective values between these simplified versions and our proposed algorithm for each Beasley problem set. Each implementation was executed for the same number of epochs as outlined in the \textit{\alg} column of Table \ref{Table:2}.

\begin{figure*}[tbp]
    \centering
    \begin{minipage}{.33\textwidth}
        \includegraphics[width=0.85\linewidth]{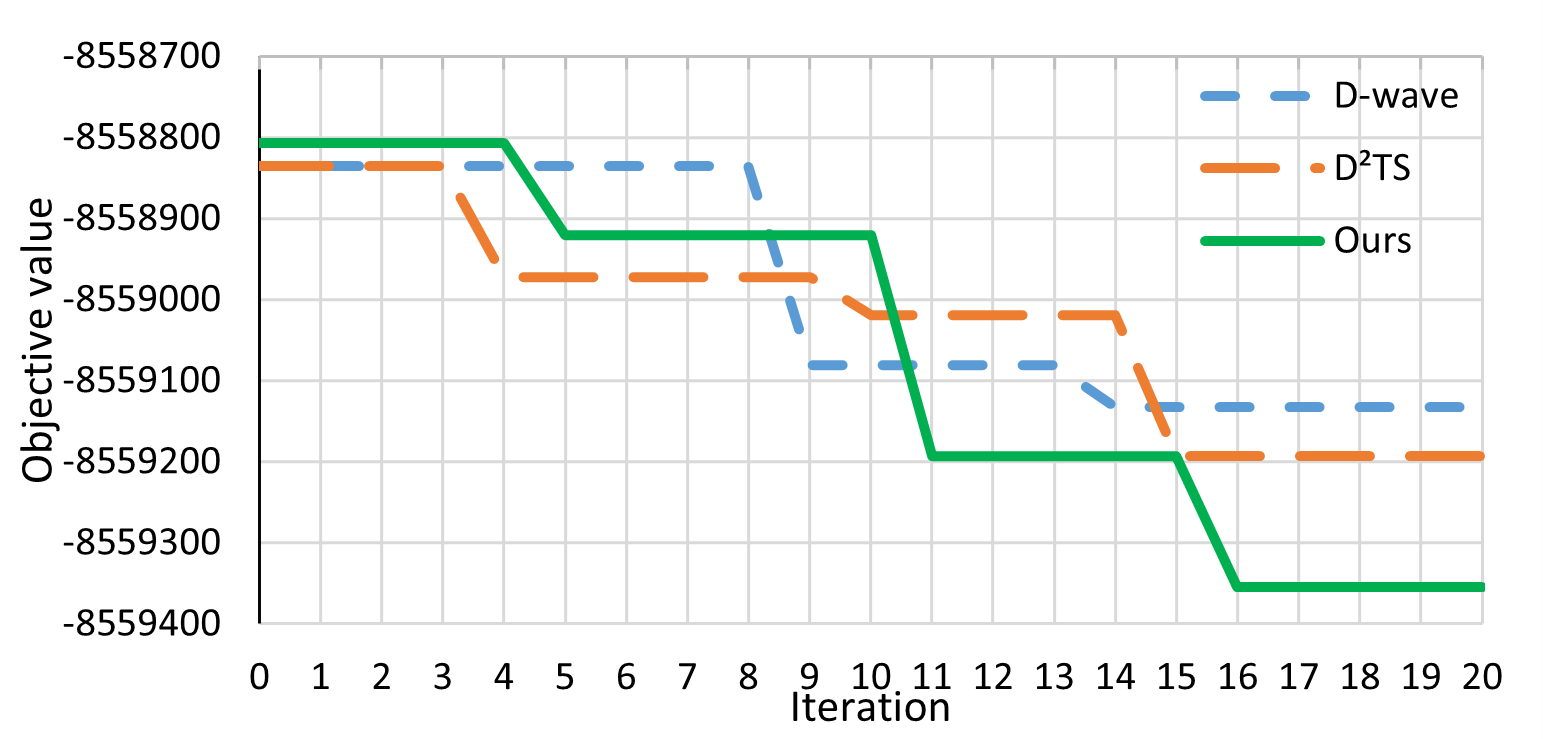}
        \centering \small \textbf{(a)} P5000\_1
    \end{minipage}\hfill
    \begin{minipage}{.33\textwidth}
        \includegraphics[width=0.85\linewidth]{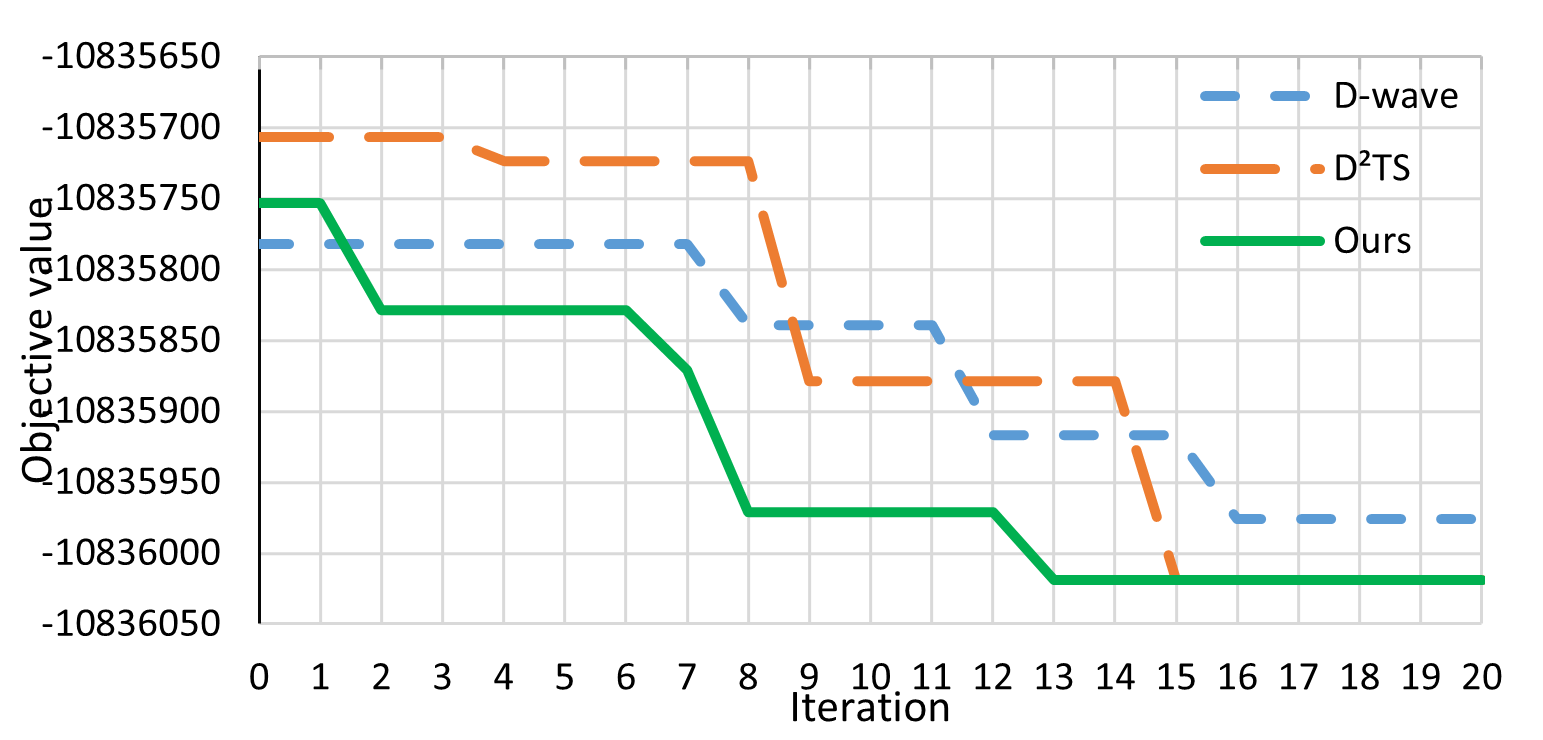}
        \centering \small \textbf{(b)} P5000\_2
    \end{minipage}\hfill
    \begin{minipage}{.33\textwidth}
        \includegraphics[width=0.85\linewidth]{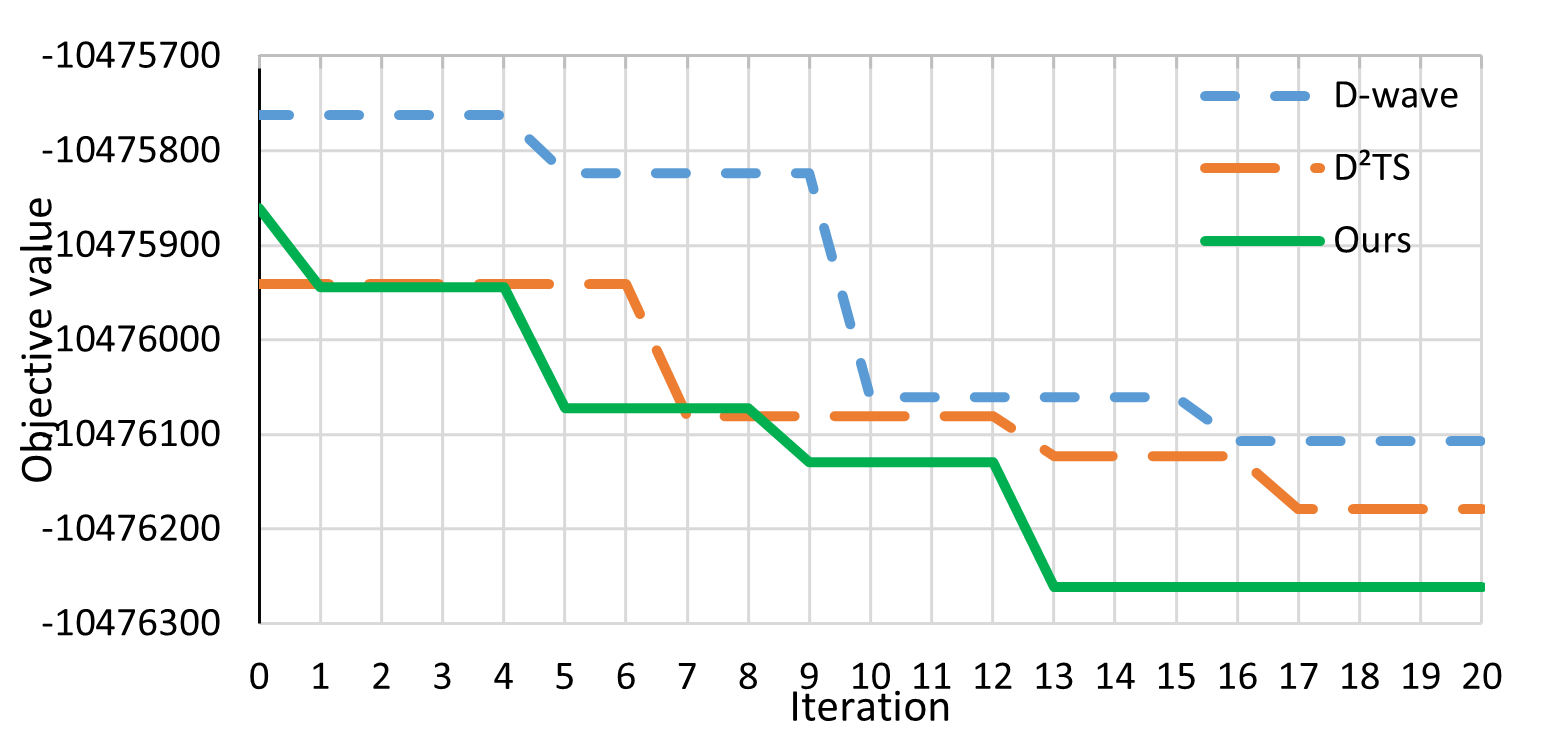}
        \centering \small \textbf{(c)} P5000\_3
    \end{minipage}
    
    \vfill 

    \caption{\small Overall comparison of three different algorithms on the problems with a size of 5000 and varying densities.}
    \label{fig:overall5}
    \vspace{-1em}
\end{figure*}

\begin{figure*}[tbp]
    \centering
    \begin{minipage}{.33\textwidth}
        \includegraphics[width=0.85\linewidth]{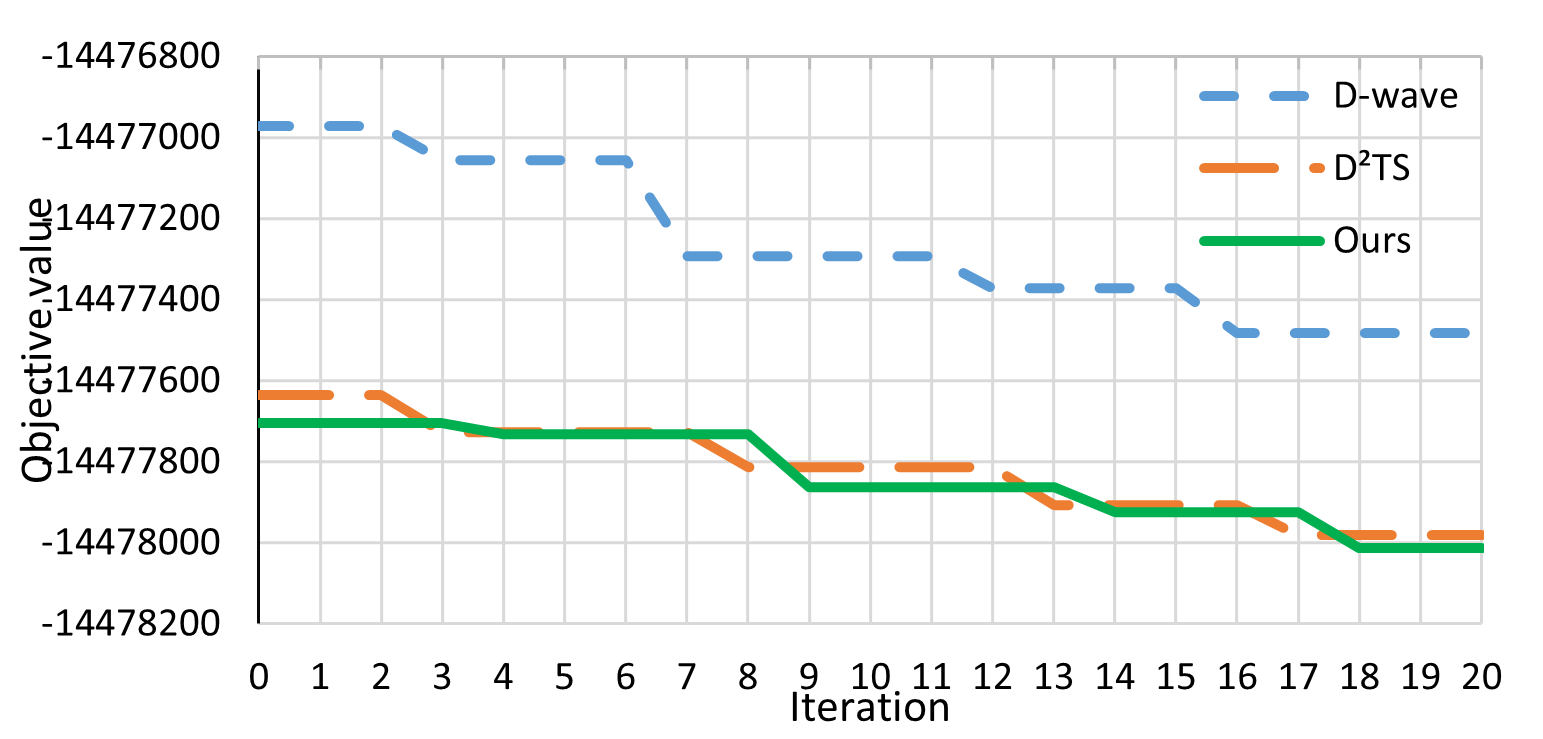}
        \centering \small \textbf{(a)} P7000\_1
    \end{minipage}\hfill
    \begin{minipage}{.33\textwidth}
        \includegraphics[width=0.85\linewidth]{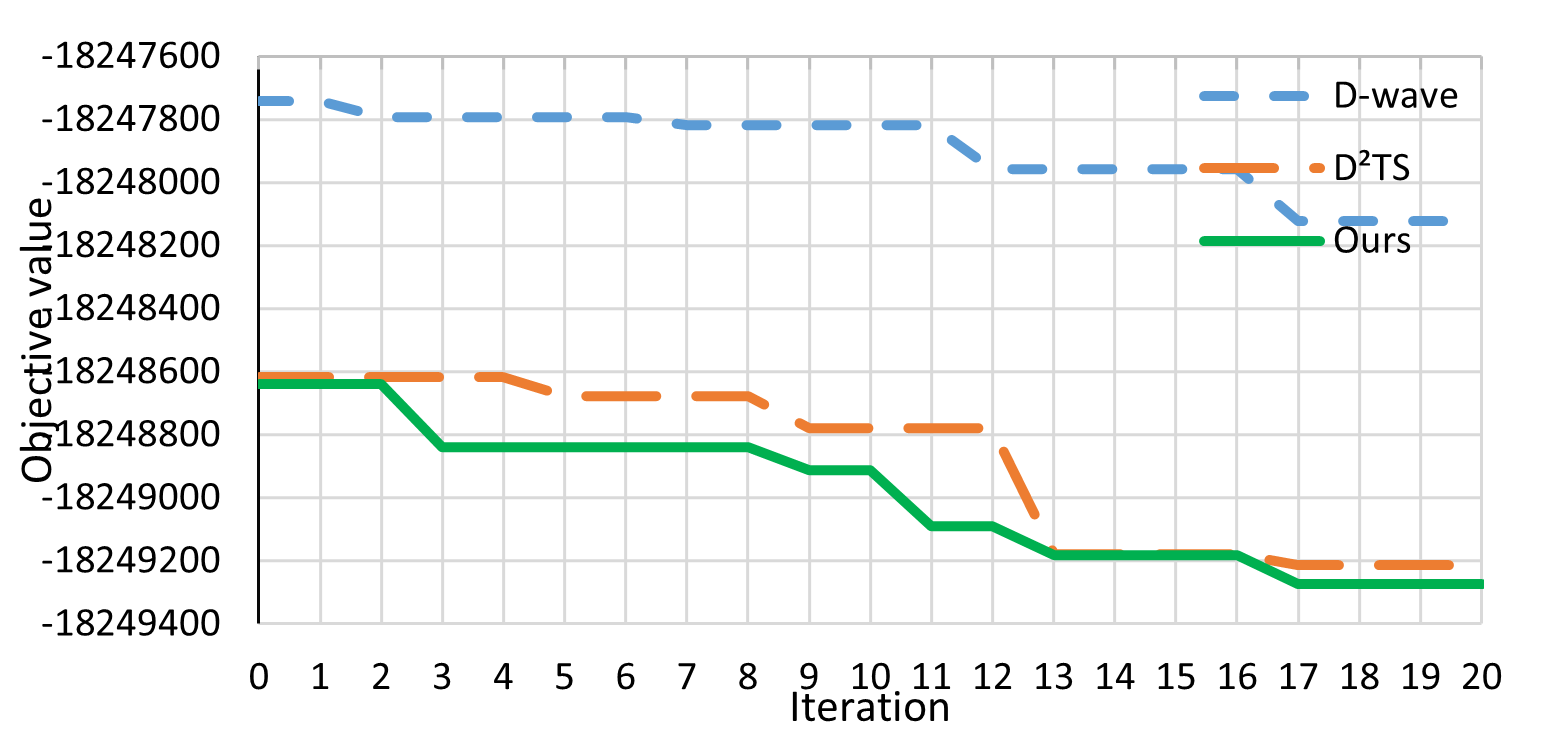}
        \centering \small \textbf{(b)} P7000\_2
    \end{minipage}\hfill
    \begin{minipage}{.33\textwidth}
        \includegraphics[width=0.85\linewidth]{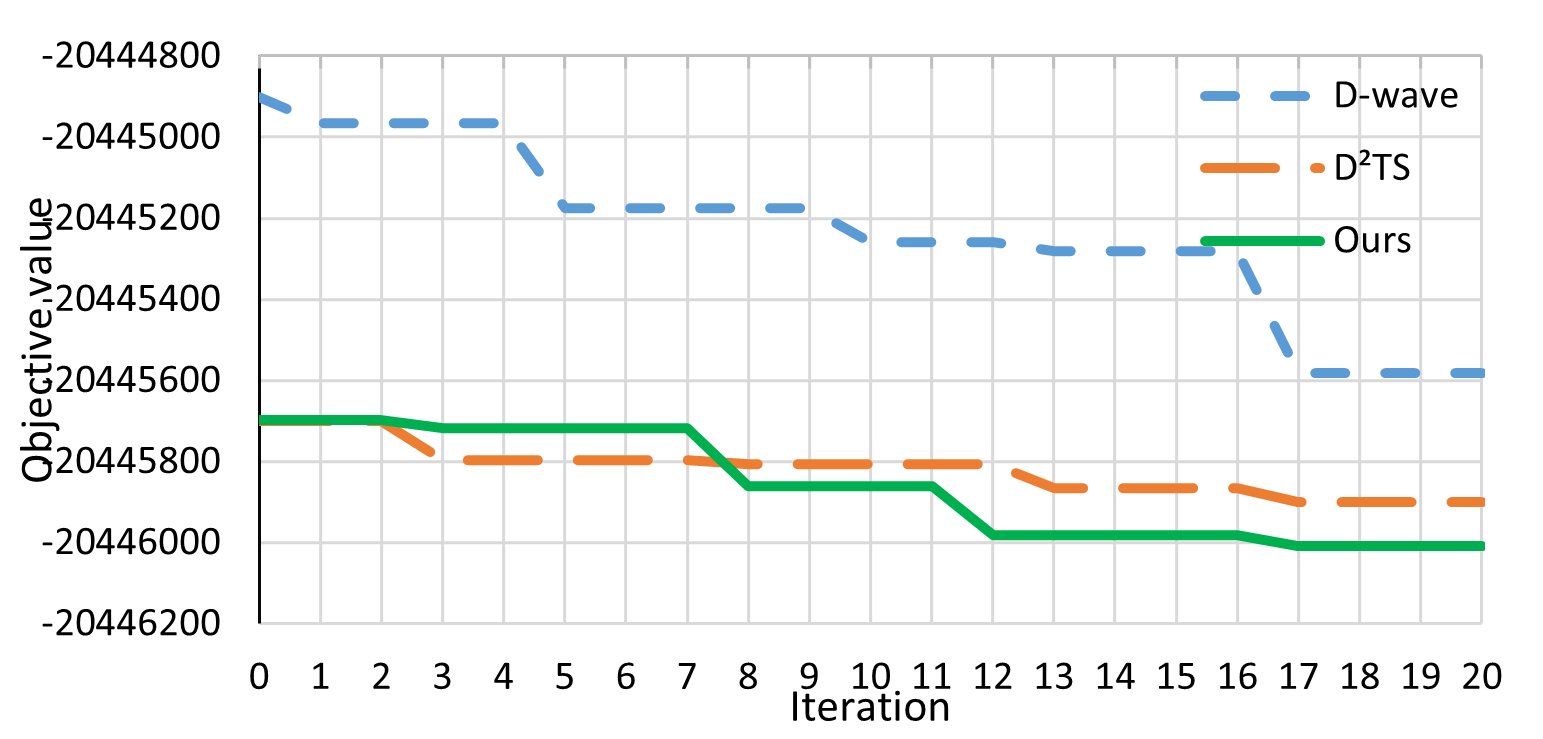}
        \centering \small \textbf{(c)} P7000\_3
    \end{minipage}
    
    \vfill 
    \caption{\small Overall comparison of three different algorithms on the problems with a size of 7000 and varying densities.}
    \label{fig:overall7_5}
    \vspace{-1em}
\end{figure*}

\begin{table}[htp]
\caption{\small Optimal objective value of the QUBO problems with the size of 2,500 and the number of iterations needed by three considered algorithms to find the best objective values}
\label{Table:2}
\centering
\vspace{-1em}
\renewcommand{\arraystretch}{0.4}
\begin{tabular}{|c|c|c|c|c|}
\hline
\textbf{Problem} & \textbf{Min Value} & \textbf{D-wave} & \textbf{D$^2$TS} & \textbf{\alg} \\
\hline
Q1 & -1515944 & 1 & 1 & 1\\
\hline
Q2 & -1471392 & 28 & 26 & 12 \\
\hline
Q3 & -1414192 & 24 & 23 & 5\\
\hline
Q4 & -1507701 & 1 & 1 & 1\\
\hline
Q5 & -1491816 & 4 & 2 & 1\\
\hline
Q6 & -1469162 & 10 & 3 & 1\\
\hline
Q7 & -1479040 & 17 & 3 & 3\\
\hline
Q8 & -1484199 & 8 & 12 & 1\\
\hline
Q9 & -1482413 & 9 & 11 & 1\\
\hline
Q10 & -1483355 & 26 & 24 & 8\\
\hline
\end{tabular}
\end{table}

\vspace{-10pt}

The comparison reveals that, except for Q1 and Q5, the No\_SM and No\_IM versions could not reach the optimal solutions. Despite the relatively small difference in objective values between these simplified heuristics and \alg, achieving the final objective value was significantly more challenging for them. For instance, in the Q2 problem, No\_IM failed to find the optimal solution within 200 epochs, while No\_SM achieved the best solution in just 25 epochs. The results indicate that incorporating control metrics in our heuristic has a slightly larger impact on efficiency than using the Ising machine alone. However, the synergistic combination of these techniques highlights the superior performance of our algorithm.

The objective values of problems with sizes of 5,000 and 7,000, obtained by the three considered algorithms over different search epochs, are shown in Figure\ref{fig:overall5} and Figure \ref{fig:overall7_5}, respectively. As illustrated in these figures, our proposed heuristic shows significant progress over the iterations, demonstrating its effectiveness in escaping local minima. This heuristic consistently finds the best solutions compared to the other two heuristics. In Table \ref{Table:3}, the success rate of different heuristics, defined as the frequency of achieving the optimal solution across multiple runs, is demonstrated. Among the hybrid heuristics that utilize both a classical optimizer and an Ising machine, our heuristic shows a superior improvement in the success rate for Beasley and Palubeckis matrices. Additionally, our hybrid heuristic outperforms even the pure classical heuristic.

\begin{table}[t]
\caption{\small Success rate of different heuristics on Beasley and Palubeckis matrices}
\vspace{-1em}
\label{Table:3}
\centering
\renewcommand{\arraystretch}{0.1}
\begin{tabular}{|c|c|c|c|c|}
\hline
\textbf{Problem} & \textbf{D$^2$TS} & \textbf{D-wave} & \textbf{\alg}  & \textbf{\cite{ref28}} \\
\hline
Q1 & 10/10 & 10/10 & 10/10 & 4/10 \\
\hline
Q2 & 10/10 & 10/10 & 10/10 & 0/10 \\
\hline
Q3 & 10/10 & 10/10 & 10/10 & 1/10 \\
\hline
Q4 & 10/10 & 10/10 & 10/10 & 10/10 \\
\hline
Q5 & 10/10 & 10/10 & 10/10 & 6/10 \\
\hline
Q6 & 10/10 & 10/10 & 10/10 & 3/10 \\
\hline
Q7 & 10/10 & 10/10 & 10/10 & 2/10 \\
\hline
Q8 & 10/10 & 10/10 & 10/10 & 5/10 \\
\hline
Q9 & 10/10 & 10/10 & 10/10 & 4/10 \\
\hline
Q10 & 10/10 & 10/10 & 10/10 & 0/10 \\
\hline
P5000\_1 & 6/10 & 3/10 & 8/10 & 0/10 \\
\hline
P5000\_2 & 7/10 & 4/10 & 7/10 & 0/10 \\
\hline
P5000\_3 & 8/10 & 4/10 & 8/10 & 1/10 \\
\hline
P5000\_4 & 4/10 & 2/10 & 6/10 & 0/10 \\
\hline
P5000\_5 & 9/10 & 7/10 & 9/10 & 0/10 \\
\hline
P7000\_1 & 4/10 & 4/10 & 7/10 & 0/10 \\
\hline
P7000\_2 & 7/10 & 5/10 & 8/10 & 0/10 \\
\hline
P7000\_3 & 3/10 & 3/10 & 6/10 & 0/10 \\
\hline
\end{tabular}
\vspace{-1em}
\end{table}

\vspace{-13pt}

\section{Conclusion}
\vspace{-2pt}
We introduced a hybrid classical processing and Ising machine algorithm designed to solve large QUBO problems. 
A fundamental aspect of our algorithm entails the specification of control parameters to navigate the search space and generate subQUBO problems that can be solved on a small Ising machine, alongside the implementation of a mutation step whose operation is overseen by the control parameters and cosine annealing. 
Through comprehensive experimentation conducted across many problem instances, our results demonstrated the efficacy and superiority of our hybrid algorithm in terms of convergence rate and solution quality, consistently surpassing established methodologies. 

\vspace{-5pt}
\section{Acknowledgments}
This work has been funded by the National Science Foundation (NSF)
under the project Expedition: (Design and Integration of Superconducting Computation for Ventures beyond Exascale Realization) project with grant number 2124453.

\bibliographystyle{unsrt}
\bibliography{myref}

\end{document}